\documentclass[runningheads]{llncs}

\usepackage{color} 
\usepackage{xspace} 
\usepackage{booktabs}
\usepackage{enumitem}
\usepackage{balance}
\usepackage{graphicx}

\usepackage{courier}
\usepackage{amsmath}
\usepackage{soul}
\usepackage{color}

\usepackage{enumitem}
\usepackage{booktabs}
\usepackage{multirow}
\usepackage{adjustbox}
\usepackage{dcolumn}

\usepackage{subcaption}
\usepackage{multirow}
\usepackage{url}
\usepackage{wrapfig}

\begin{document}
	
	\title{A Reproducibility Study of Goldilocks: Just-Right Tuning of BERT for TAR}
	
	\author{Xinyu Mao\inst{1}\orcidID{0000-0001-6357-2311} \and
		Bevan Koopman\inst{1,2}\orcidID{0000-0001-5577-3391} \and
		Guido Zuccon\inst{1}\orcidID{0000-0003-0271-5563}}
	
	\institute{The University of Queensland, St. Lucia, Australia \\
		\email{\{xinyu.mao, g.zuccon\}@uq.edu.au}
		\and 
		CSIRO, Brisbane, Australia \\
		\email{bevan.koopman@csiro.au}}

	\maketitle              

\begin{abstract}
Screening documents is a tedious and time-consuming aspect of high-recall retrieval tasks, such as compiling a systematic literature review, where the goal is to identify all relevant documents for a topic. To help streamline this process, many Technology-Assisted Review (TAR) methods leverage active learning techniques to reduce the number of documents requiring review. BERT-based models have shown high effectiveness in text classification, leading to interest in their potential use in TAR workflows. 
In this paper, we investigate recent work that examined the impact of further pre-training epochs on the effectiveness and efficiency of a BERT-based active learning pipeline. We first report that we could replicate the original experiments on two specific TAR datasets, confirming some of the findings: importantly, that further pre-training is critical to high effectiveness, but requires attention in terms of selecting the correct training epoch. We then investigate the generalisability of the pipeline on a different TAR task, that of medical systematic reviews. In this context, we show that there is no need for further pre-training if a domain-specific BERT backbone is used within the active learning pipeline. This finding provides practical implications for using the studied active learning pipeline within domain-specific TAR tasks.

	
\vspace{-12pt}
\end{abstract}
	
\section{Introduction}

Review tasks in professional domains often require screening of a large number of documents to ensure all evidence about a review topic is identified. This task is often associated with high recall retrieval (HRR); examples of such tasks include the compilation of systematic literature reviews, legal eDiscovery, and prior-art finding in patent applications~\cite{grossman2011technology,lupu2014domain,MICHELSON2019100443}. The manual screening of a large number of candidate documents can be time-consuming and resource-intensive. To reduce the number of documents needing manual review, automated methods are used to identify the relevant documents in a given set, with the aim of achieving a targeted recall: a process known as technology-assisted review (TAR). 

Pre-trained language models such as BERT~\cite{devlin2018bert}, T5~\cite{raffel2020exploring} and GPT~\cite{radford2019language} have exhibited state-of-the-art effectiveness in tasks such as general-domain search~\cite{tonellotto2022lecture,lin2021pretrained}, question answering~\cite{karpukhin2020dense}, or text summarization~\cite{liu2019text}. These language models follow the transformer architecture~\cite{vaswani2017attention} and are able to model word semantics by performing a pre-training step, e.g., masked language modelling (MLM) \cite{vaswani2017attention}.


 In this work, we focus on reproducing the TAR pipeline of Yang et al. that exploits pre-trained language models~\cite{yang2022goldilocks}. Pre-trained language models~\cite{devlin2018bert,raffel2020exploring,radford2019language} have exhibited state-of-the-art effectiveness across several tasks~\cite{tonellotto2022lecture,lin2021pretrained,karpukhin2020dense,liu2019text}. In the pipeline of Yang et al, the TAR task is modelled as a binary classification problem on a large dataset with categories and their pipeline uses active learning to continuously fine-tune BERT classifiers. A key aspect of the pipeline is the additional pre-training of the BERT backbone to the document collection of the TAR task, which is critical to obtain effectiveness improvements compared to a baseline logistic regression approach.

 \begin{wrapfigure}{r}{0.5\textwidth}
 	\centering
 	\vspace{-20pt}
 	\includegraphics[width=0.5\textwidth]{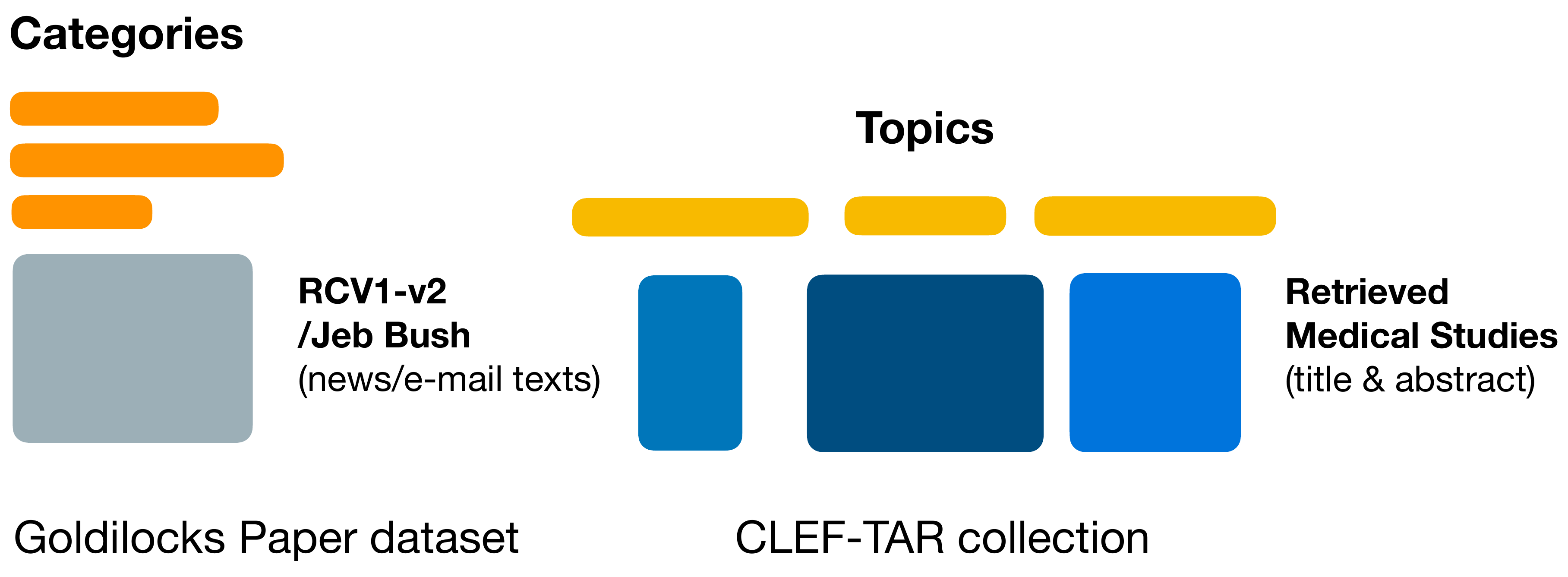}
 	\vspace{-10pt}
 	\caption{Difference in task configuration between datasets.}
 	\label{fig:1}
 	\vspace{-20pt}
 \end{wrapfigure}
 In this paper, we aim to reproduce and verify the findings of the original study and investigate whether the original conclusions generalise to other TAR tasks, and in particular that of a medical systematic review literature search ~\cite{kanoulas2017clef,kanoulas2018clef,kanoulas2019clef}. The study of the pipeline's generalisability is not trivial because the TAR tasks considered in the original work crucially differ from those in medical systematic reviews for the following aspects, which are also visualised in Fig.~ \ref{fig:1}:

\begin{itemize}[leftmargin=8pt]
	\item \textbf{Domain:} They contain biomedical literature and thus might require a more domain-specific approach.
	\item  \textbf{Proposed task:} They are designed to evaluate ranking instead of classification, with the goal of ranking the set of documents associated with the topic in decreasing order of relevance.
	\item  \textbf{Dataset composition:} They are composed of topics, each of which represents a systematic review, and a set of corresponding documents that are retrieved by a Boolean query and provided with only the title and abstract. TAR is performed separately and independently for each topic: this means that the TAR task is performed on different sources of documents as reference for each topic.
	This is unlike the datasets in the original work where the TAR task was performed once for each dataset. This is a crucial difference because (1) the size of the document set on which TAR is performed largely differs between the original and our setup, and (2) the nature of the classes assigned largely differ -- very distinct classes in the original work vs. inclusion/exclusion classes related to topical relevance in our setup.
\end{itemize}

Our investigation provides insights into the reproducibility and generalisability of the original work by Yang et al., in particular in the context of applying their BERT-based active learning pipeline to the context of medical systematic review creation. Importantly, we identify a crucial difference in the experimental findings that have implications for the practical use of the pipeline within medical systematic review settings. Code, results and supporting material that could not be included in the paper due to space limitations can be found at  \url{https://github.com/ielab/goldilocks-reproduce}.

\vspace{-10pt}
\section{Goldilocks: Just-Right Tuning of BERT}
\vspace{-10pt}

 
%
%
\subsection{Overall Active Learning Pipeline} \label{tar workflow}
\begin{figure}[t]
	\includegraphics[width=0.85\columnwidth]{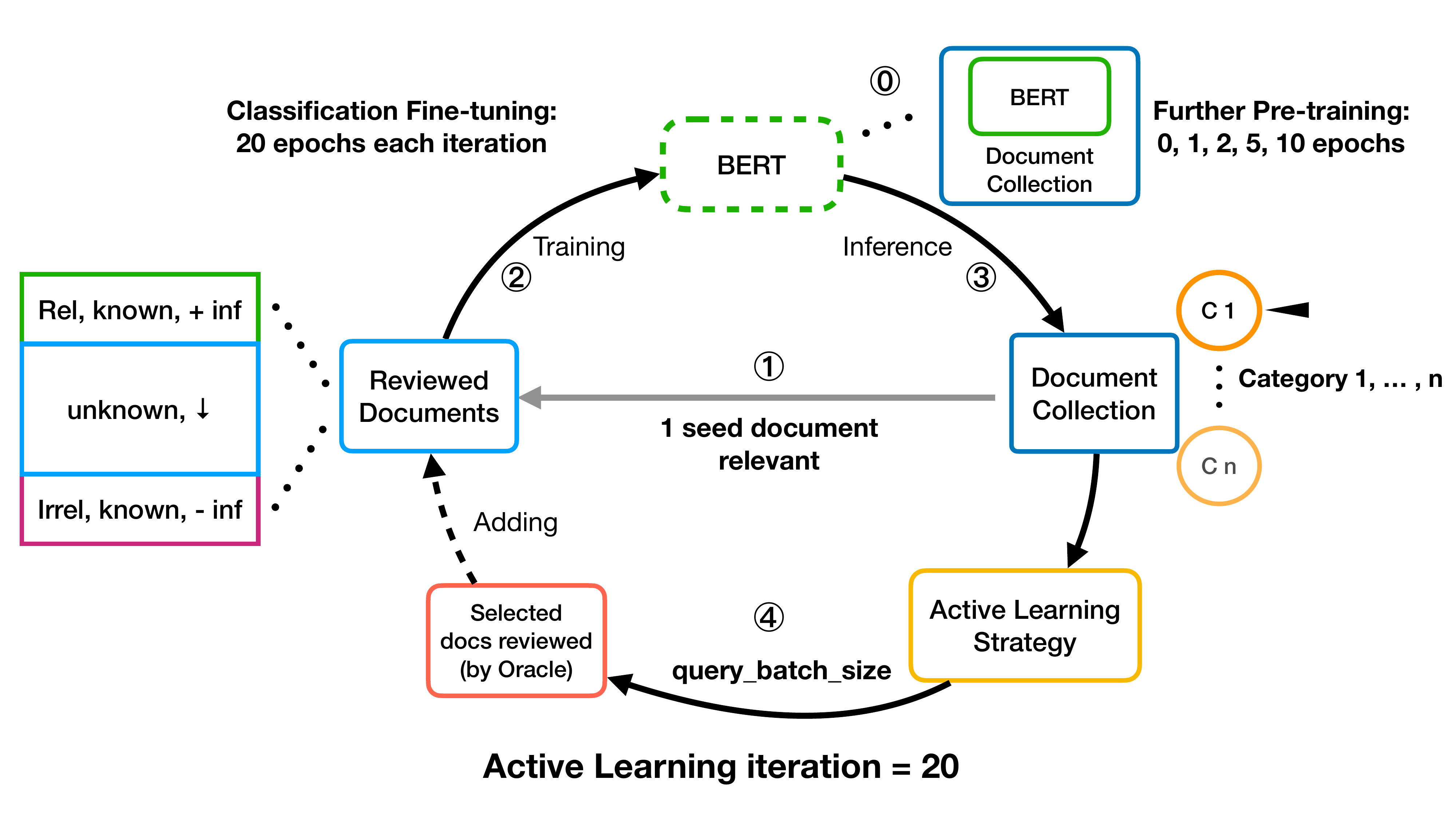}
	\vspace{-10pt}
	\caption{The TAR workflow using active learning with BERT used by the Goldilocks paper.} 
	\vspace{-20pt}
	\label{fig:2}
\end{figure}

Yang et al. have proposed a pipeline that uses a BERT classifier to undertake a TAR task within an active learning workflow~\cite{yang2022goldilocks}. The pipeline is visualised in Fig. \ref{fig:2}, and it consists of two key components: (1) Further Pre-training and; (2) Fine-tuning.

\textbf{(1) Further Pre-training.} Prior to the active learning process, a BERT pre-trained model undergoes additional pre-training using MLM on the entire target data. A key aspect of the original study was to investigate the effect that varying the number of epochs used in the further pre-training task has on the final model's effectiveness. 

\textbf{(2) Fine-tuning.} The fine-tuning regime followed four steps:

\begin{enumerate}[label=\roman*,labelindent=0pt]
	\item \textit{Initiating with seed document.} A seed document for a specific target category is specified to initiate the active learning process. A seed document is a document that is relevant to the category.
	
	\item \textit{Classification fine-tuning.} BERT is fine-tuned with all the reviewed documents in the training set for a fixed number of epochs. The original paper demonstrated that fine-tuning using only newly labelled documents has worse performance. At the start, the reviewed documents only consist of the seed document. In subsequent iterations, more reviewed documents are added. 
	\item \textit{Scoring all remaining documents.} The fine-tuned model is used to score the remaining documents in the dataset. The scores are used by an active learning strategy (relevance feedback \cite{lewis1995sequential} or uncertainty sampling \cite{salton1990improving}) to select informative document samples to be reviewed/labelled by reviewers (known as the oracle). Already reviewed documents are excluded from selection.
	\item \textit{Querying and labelling new documents and updating the training set.} The newly reviewed documents are added to the training set. 
\end{enumerate}

Steps ii, iii, and iv are repeated iteratively until all iterations are completed.

\subsection{Key Findings from the Original Study}
\vspace{-6pt}
The key findings from the original study that we aim to reproduce are:
\begin{itemize}[leftmargin=8pt]
	
	\item Key to the effectiveness of the proposed pipeline is the existence of a Goldilocks (or 'just-right') epoch of further pre-training. This refers to the further pre-training step outlined in Sec.~\ref{tar workflow}. The correct Goldilocks epoch depends on the dataset and task characteristics. An incorrect setting makes the BERT classifier underperform a baseline logistic regression, even on in-domain tasks.
	
	
	\item Further pre-training does not solve the issue of domain mismatch. Domain mismatch occurs when the data used for the pre-training does not match the data used at inference. Further pre-training on the target dataset is expected to address domain mismatch, but this is not the case when the target is out-of-domain, even when the correct Goldilocks epoch is identified.
	
	
	\item The computational costs associated with the proposed pipeline discourage its use because effectiveness improvements are only marginal compared to the logistic regression baseline.
\end{itemize}

\vspace{-10pt}
\section{Experiment Setup}
\vspace{-10pt}
We devise a set of experiments to address the following research questions (RQs):
\begin{description}[leftmargin=4pt]
	\item[RQ1]   \textit{To what extent the original results can be reproduced?}
	
	\item[RQ2]  \textit{How well do the original findings generalise to the task of screening in medical systematic review creation?}
	
	\item[RQ3]  \textit{Is there a domain-mismatch problem in the task of screening in medical systematic review creation, and can this be mitigated by adopting a BERT backbone specifically designed for the biomedical domain?}
	
\end{description}

\vspace{-17pt}
\subsection{Datasets}
\vspace{-8pt}
We use three groups datasets for our experiment: RCV1-v2 \cite{lewis2004rcv1}, Jeb Bush \cite{roegiest2015trec,grossman2016trec}, both used in the original study \cite{yang2022goldilocks}, and the CLEF TAR collections \cite{kanoulas2017clef,kanoulas2018clef,kanoulas2019clef}. The CLEF TAR consists of several datasets.

	\textbf{RCV1-v2 \cite{lewis2004rcv1}:} consists of newswire articles, and as such was regarded as in-domain for BERT by Yang et al..
	The original study considered 658 categories, each containing no less than 25 documents, and selected 45 categories according to prevalence (rare, medium, common) and difficulty (hard, medium, easy) following previous works \cite{yang2018retrieval,yang2021minimizing}. Then they downsampled to 160,833 documents (20\%) for computational efficiency. We obtained a table of these 45 categories from Yang et al., and downsampled by ourselves with the same rate. 
	
	
	\textbf{Jeb Bush \cite{roegiest2015trec,grossman2016trec}:} consists of email texts related to specific local issues of a constituent, and because of this Yang et al. regarded this dataset as out-of-domain for BERT. The dataset contains 44 topics and 290,099 emails. We  followed Yang et al. in de-duplicating these emails by comparing the md5 hash string of each email text, to obtain 274,124 unique emails. The emails are further randomly downsampled to 137,062 (50\%) following the original study; this makes the dataset comparable in size to the downsampled RCV1-v2 dataset.


   \textbf{CLEF TAR collections \cite{kanoulas2017clef,kanoulas2018clef,kanoulas2019clef}:} comprise six datasets -- CLEF17 test (30 topics, total 117,557 documents), CLEF17 train (20 topics, total 149,404 documents), CLEF18 test (30 topics, total 218,484 documents), CLEF19 dta test (8 topics, total 30,521 documents), which include diagnostic test accuracy reviews, and CLEF19 intervention test (20 topics, total 41,996 documents), CLEF19 intervention train (20 topics, total 31,639 documents), which include intervention reviews. Each dataset contains a set of systematic review topics, along with documents (title and abstract of studies considered for inclusion in the systematic review) and inclusion/exclusion judgements. Although originally proposed to evaluate a ranking task (screening prioritisation), we adapt the datasets for classification (screening) similar to the RCV1-v2 and Jeb Bush datasets, so that we treat each topic as a separate classification task. We consider only the original splits of these datasets, which for each year include train and test splits, but we do not perform training on the portions labelled ``train'': instead we used them all for evaluation. This is because classifiers in these tasks are not trained across different reviews -- they are instead trained ``on-the-spot'' on the specific systematic review under consideration, in an active learning setting.
   
   We report further details on the datasets along with information on their pre-processing in the online repository.

\vspace{-10pt}
\subsection{Considered Methods: Baseline and BERT Model}
\vspace{-10pt}
We follow the original study in using a logistic regression (LR) classifier as the baseline model for comparison \cite{yang2022goldilocks}; as in the original study, we also source the implementation from \texttt{scikit-learn}\footnote{https://scikit-learn.org/stable/}. To replicate their BERT-based classifier, we acquire the common \texttt{bert-base-cased}\footnote{https://huggingface.co/bert-base-cased} backbone from Hugging Face.
The active learning implementation is based on the \texttt{libact}\footnote{https://github.com/ntucllab/libact} library. We requested code from the original authors and obtained the BERT part, while we adapted related code for the baseline part, including tokenisation and training.

%

\vspace{-10pt}
\subsection{Evaluation Metrics}
\vspace{-10pt}
We report the same evaluation metrics as in the original study, including R-Precision, the uniform and expensive training cost \cite{yang2022goldilocks}. 

Recall-Precision (R-Precision) calculates the proportion of relevant documents retrieved among the top R retrieved documents, where R is the total number of relevant documents (total recall) for a given category or topic: a recall target of 80\% is common in eDiscovery; 95\% in systematic reviews.

The review cost is characterised by the cost structure $(\alpha_{p}, \alpha_{n}, \beta_{p}, \beta_{n})$, where: $\alpha_{p}$ and $\alpha_{n}$ are the unit cost of training a classifier on reviewed relevant and irrelevant documents respectively during the first phase; $\beta_{p}$ and $\beta_{n}$ are the unit cost of reviewing relevant and irrelevant unreviewed documents needed to attain the target recall post active learning training, i.e., in the second phase.
The review cost is then computed as the cumulative product of the cost structure coefficients and the corresponding document numbers: $\alpha_{p} t_{p} + \alpha_{n}t_{n} + \beta_{p}m_{p} + \beta_{n}m_{n}$, where $t_{p}$ and $t_{n}$ are the number of relevant and irrelevant documents reviewed for training the classier at the first phase, while $m_{p}$ and $m_{n}$ are the number of remaining relevant and irrelevant documents to obtain the target recall in the second phase. Following the original study, we report the minimal total cost observed over the 20 active learning iterations and set the target recall to 80\% \cite{yang2022goldilocks}. We also consider two cost structures: (1,1,1,1) referred to as uniform, and (10,10,1,1) referred to as expensive (a cost of 10 is assigned to the training process, leaving the rest unchanged).

We report the statistical significance of the differences between methods using a paired t-test ($\alpha=0.05$) with Bonferroni correction, as in the original study.

\vspace{-10pt}
\subsection{Parameters Setting and Experiment Environment}
\vspace{-10pt}
For executing the active learning process  we use 3 HPC server nodes, each equipped with 3 NVIDIA A100 GPUs with 80GB of memory per GPU for tokenization, further pre-training, fine-tuning, and inference. For the baselines using logistic regression, we run on a 48-core AMD CPU. 
The hyperparameters for BERT and logistic regression are set based on the original paper; further details are provided in the online documentation.


For the active learning pipeline, we use the same two sampling strategies as in the original paper, namely, relevance feedback and uncertainty sampling, with a query batch size of 200 for RCV1-v2 and Jeb Bush. However, for the CLEF collections, due to the large variation in the total number of documents for each topic, we set a moderate batch size value of 25; we did so informed by a previous work that executed activate learning for systematic reviews and considered a batch of 25 candidate documents as reasonable for the workflow \cite{singh2018improving}.


The analysis was performed based on our understanding of the metrics, as there were no results files from the original authors to directly test and compare. All of the results and analyses presented in this study are based on our own experiments using the materials described above.

\section{Results}

\subsection{RQ1: Is the Goldilocks Finding Replicable?}

\begin{table}[!t]
	\newsavebox{\CBox} 
	\def\textBF#1{\sbox\CBox{#1}\resizebox{\wd\CBox}{\ht\CBox}{\textbf{#1}}}
	\newcommand{\minus}{\scalebox{0.8}[1.0]{$-$}}
	\scriptsize
	\centering
	\caption{Results for RCV1-v2 and Jeb Bush dataset. In brackets for R-Precision: percentage difference between our results and those of the original study.  Both uniform cost (Uni. Cost) and expensive training cost (Exp. Train.) values are presented as the relative cost difference between LR and BERT; the lower the value, the better. 
		\textsuperscript{$\ast$} indicates a statistically significant difference w.r.t. the baseline LR.}
	\label{table:1}
	\begin{adjustbox}{width=1.0\textwidth, center=\textwidth}
		\begin{tabular}{@{}cc|cc|rr|rrr@{}}
			\toprule
			\multirow{2}{*}{Dataset}                                                    & \multirow{2}{*}{\begin{tabular}[c]{@{}c@{}}Further Pre-training \\ Epoch\end{tabular}} & \multicolumn{2}{c|}{R-Precision ($\uparrow$)}               & \multicolumn{2}{r|}{Uni. Cost ($\downarrow$)} & \multicolumn{2}{r}{Exp. Train. ($\downarrow$)} \\
			&                                                                                        & Relevance              & Uncertainty           & Rel.           & Unc.          & Rel.           & Unc.           \\ \midrule
			\multirow{6}{*}{\begin{tabular}[c]{@{}c@{}}In-domain\\ RCV1-v2\end{tabular}}   
			& LR                                                                                     & \textBF{0.754} (\minus4.32\%)   & \textBF{0.733} (\minus3.51\%)  & \textBF{1.000}          & \textBF{1.000}         & \textBF{1.000}          & \textBF{1.000}          \\
			& 0                                                                                      &0.688 (\minus8.51\%)  &0.740 (\minus2.08\%)  &\textsuperscript{$\ast$}1.720 &\textsuperscript{$\ast$}2.059 &  \textsuperscript{$\ast$}1.386 &  \textsuperscript{$\ast$}1.516         \\
			& 1                                                                                      & 0.759 (+0.20\%)  & 0.810 (+5.44\%)   & 1.004 & 1.091 & 0.969 & 1.014         \\
			& 2                                                                                      &0.771 (+1.52\%)  & 0.822 (+7.32\%)  & 0.935 & 1.053 & 0.935 & 0.987          \\
			& 5                                                                                      & 0.770 (+1.85\%)  & 0.819 (+4.46\%)  & 0.997 & 0.991 & 0.916 & \textBF{0.946}         \\
			& 10                                                                                     &\textBF{0.785} (+2.72\%)  & \textBF{0.838} (+9.52\%)  & \textBF{0.823} & \textBF{0.967} & \textBF{0.857} & 0.952          \\ \midrule
			\multirow{6}{*}{\begin{tabular}[c]{@{}c@{}}Off-domain\\ Jeb Bush\end{tabular}} & LR                                                                                     & \textBF{0.892} (\minus1.34\%)   & \textBF{0.866} (+1.04\%)  & \textBF{1.000}          & \textBF{1.000}         & \textBF{1.000}          & \textBF{1.000}          \\
			& 0                                                                                      & \textsuperscript{$\ast$}0.528 (\minus27.07\%) &\textsuperscript{$\ast$}0.487 (\minus32.22\%) &\textsuperscript{$\ast$}6.471 &\textsuperscript{$\ast$}4.545 &\textsuperscript{$\ast$}3.609 &\textsuperscript{$\ast$}2.890            \\
			& 1                                                                                      & \textsuperscript{$\ast$}0.635 (\minus21.64\%) & \textsuperscript{$\ast$}0.569 (\minus30.21\%) & \textsuperscript{$\ast$}4.223 & \textsuperscript{$\ast$}3.239 & \textsuperscript{$\ast$}2.736 & \textsuperscript{$\ast$}2.317        \\
			& 2                                                                                      & \textsuperscript{$\ast$}0.678 (\minus16.54\%) & \textsuperscript{$\ast$}0.602 (\minus25.53\%) & \textsuperscript{$\ast$}3.439 & 2.691 & \textsuperscript{$\ast$}2.410 & \textsuperscript{$\ast$}1.976          \\
			& 5                                                                                      & \textsuperscript{$\ast$}\textBF{0.706} (\minus12.90\%) & \textsuperscript{$\ast$}\textBF{0.632} (\minus22.30\%) & \textsuperscript{$\ast$}\textBF{2.748} & \textBF{2.509} & \textsuperscript{$\ast$}\textBF{2.079} & \textsuperscript{$\ast$}\textBF{1.858}          \\
			& 10                                                                                     & \textsuperscript{$\ast$}0.701 (\minus12.95\%) & \textsuperscript{$\ast$}0.600 (\minus26.35\%) & \textsuperscript{$\ast$}3.049 & 2.867 & \textsuperscript{$\ast$}2.199 & \textsuperscript{$\ast$}1.897          \\ \bottomrule
		\end{tabular}
	\end{adjustbox}
\vspace{-18pt}
\end{table}

\subsubsection{Further Pre-training.}

Results related to RQ1 are reported in Table \ref{table:1}, where R-Precision values are obtained from the final active learning iteration of all categories in the two datasets. While the original study does not mention this being the case, it is reasonable to do so because of how they recorded the values: the R-Precision increases as the reviewed relevant documents are fixed and potentially more relevant documents are prioritised at the top position of the recording in next iterations.


For RCV1-v2, the in-domain dataset, we achieved results that were largely on par or superior to those reported in the original study: the original values of R-Precision are shown in brackets in the table, along with the percentage change of our results w.r.t. the original. We find better results than the original when further pre-training occurs, regardless of the number of epochs. We instead found lower values than what was originally reported for the baseline and for the BERT classifier when no further pre-training is performed.
With respect to the key findings, we identify that additional pre-training benefits the BERT model. In particular, the BERT model further pre-trained with 10 epochs showed the largest improvement in both R-Precision and uniform training cost. However, when uncertainty sampling is used under the expensive training cost, it showed a higher cost in terms of reviewing documents compared to the BERT model pre-trained for 5 epochs.

From our replication results for this dataset, we confirm the existence of a Goldilocks epoch, with which BERT can outperform the baseline model across all the metrics reported. In our case, this was found to be 10, contrasting with the original study (in which it was 5).
Additionally, we noticed inconsistent performance of the active learning strategies on R-Precision and the two review costs. When comparing the top result under each strategy, uncertainty sampling outperformed relevance feedback on R-Precision but fell short on both uniform training cost and expensive training cost. While we were unable to replicate the original paper's precise values or their Goldilocks epoch as 5, which yielded optimal results in most settings according to their findings, our analysis offers complementary insights into the TAR workflow.

For Jeb Bush, the out-of-domain dataset, we find significant differences between the LR baseline and BERT. We also find that all our BERT results are sensibly lower than those in the original study, while for LR we only find a small difference (-1.34\%) in R-Precision. Analysing our results, we observe that LR achieves high R-Precision values, and does not exhibit poor effectiveness in the out-of-domain dataset. Conversely, BERT shows large, significant losses compared to LR. For BERT, using 5 epochs for further pre-training provides the best results across all metrics -- this is the Goldilocks epoch for this dataset. However, even at this Goldilocks epoch, BERT significantly trails behind LR in R-Precision, exhibiting an approximately 20\% decrease across both active learning strategies. 

In this case, our findings echo those of the original study: BERT, regardless of the extent of further pre-training, struggles to be an effective classifier on the Jeb Bush dataset. However, we reserve judgment on whether this is caused by the out-of-domain nature of the dataset later in our investigation.

\subsubsection{The Goldilocks Epoch Varies across Categories.}

The original study examined the impact of task characteristics on the Goldilocks epoch by examining the difficulty of categories in the RCV1-v2 dataset. We used the corresponding categories provided by the original authors, and we found the difficulty level is related to the number of relevant documents: 'Hard' means categories with less than 2000 relevant documents; 'Medium' refers to those with more than 2000 but fewer than 8000 documents, while 'Easy' denotes categories with more than 8000 relevant documents. Furthermore, each difficulty level is subdivided into three 'Prevalence' bins with further cut-off of relevant document numbers inside its range. As a result, each bin contains five distinct categories.
 Table \ref{table:2} presents our reproduced results, showing the averaged relative cost differences compared to the baseline model for each category bin under the expensive training cost structure. Contrary to the original study, our results indicate higher values for some categories while having higher review costs compared to the baseline, marked with an upper arrow in the table. Since each bin only has five runs under each further pre-training epoch setting, we did not conduct statistical tests (neither did the original study).
Our results suggest that the BERT model is generally less effective in challenging categories for both relevance feedback and uncertainty sampling. Specifically, BERT is less effective in hard categories than in easy ones for relevance feedback, and it is not better than LR even with further pre-training in this case. In contrast, for uncertainty sampling, BERT provides an improvement on additional common prevalence under hard, on top of saving numbers of documents reviewed for training on all other easier categories. These findings suggest that a linear model may be more suitable for these scenarios, and we did not perform any special feature treatments to achieve these results. It is also important to note that we only randomly sampled the same number of documents as in the original study, so our results may differ due to the potentially different documents selected.

Our findings also confirm that the trend of the Goldilocks epoch varies based on the difficulty and prevalence bins. However, our results do not reproduce the exact results of the original study. For example, we found that 2 epochs work best for the hard-medium bin instead of no further pre-training, and 1 epoch is the Goldilocks epoch for hard-rare, while the original study reported 5 epochs. These discrepancies suggest that the trend of Goldilocks is not dependent on the dataset and that it is difficult to determine a consistent Goldilocks epoch for further pre-training BERT.

\begin{table}[!t]
	\def\textBF#1{\sbox\CBox{#1}\resizebox{\wd\CBox}{\ht\CBox}{\textbf{#1}}}
	\scriptsize
	\centering
	\caption{Expensive Training (Exp. Train.) review costs for RCV1-v2 categories grouped in bins under the difficulty hierarchy. Values are the relative cost difference between the corresponding BERT models with different further pre-training epochs and the baseline LR model. Each bin under Prevalence contains 5 categories. The results are averaged over the five categories in each bin. $\uparrow$ shows results larger than both the original result and 1.}
	\begin{adjustbox}{width=1.0\textwidth, center=\textwidth}
		\begin{tabular}{@{}cc|ccccc|ccccc@{}}
			\toprule
			&  & \multicolumn{5}{c|}{Relevance} & \multicolumn{5}{c}{Uncertainty} \\
			Difficulty & Prevalence & 0 & 1 & 2 & 5 & 10 & 0 & 1 & 2 & 5 & 10 \\ \midrule
			\multirow{3}{*}{Hard} & Rare & $\uparrow$2.254 & $\uparrow$1.279 & $\uparrow$1.283 & $\uparrow$1.428 & 1.368 & $\uparrow$2.466 & $\uparrow$1.407 & $\uparrow$1.639 & $\uparrow$1.482 & $\uparrow$1.509 \\
			& Medium & $\uparrow$2.229 & $\uparrow$1.982 & $\uparrow$2.206 & $\uparrow$2.326 & 1.592 & $\uparrow$1.893 & $\uparrow$1.289 & $\uparrow$1.204 & $\uparrow$1.422 & $\uparrow$1.294 \\
			& Common & $\uparrow$1.949 & $\uparrow$1.462 & $\uparrow$1.519 & $\uparrow$1.292 & 1.381 & $\uparrow$1.050 & 0.734 & 0.729 & 0.655 & 0.883 \\ \midrule
			\multirow{3}{*}{Medium} & Rare & $\uparrow$1.339 & 0.814 & 0.799 & 0.706 & 0.728 & $\uparrow$1.557 & 0.937 & 0.879 & 0.866 & 0.814 \\
			& Medium & $\uparrow$1.600 & 1.162 & 0.947 & 0.913 & 0.945 & $\uparrow$1.422 & 0.931 & 0.882 & 0.757 & 0.804 \\
			& Common & $\uparrow$1.608 & $\uparrow$1.170 & 0.823 & 0.711 & 0.694 & $\uparrow$1.003 & 0.799 & 0.767 & 0.699 & 0.652 \\ \midrule
			\multirow{3}{*}{Easy} & Rare & 1.051 & 0.650 & 0.678 & 0.699 & 0.673 & $\uparrow$1.427 & 0.975 & 0.855 & 0.834 & 0.858 \\
			& Medium & 1.348 & 0.881 & 0.894 & 0.822 & 0.862 & $\uparrow$1.513 & $\uparrow$1.038 & $\uparrow$1.031 & 0.965 & 0.976 \\
			& Common & 0.756 & 0.694 & 0.610 & 0.543 & 0.505 & $\uparrow$1.078 & 0.898 & 0.886 & 0.837 & 0.833 \\ \bottomrule
		\end{tabular}
	\end{adjustbox}
	\label{table:2}
	\vspace{-15pt}
\end{table}

\subsubsection{Run-Time Analysis.}


In the original study, run-time was the only efficiency metric used. In our experiment, we observed an average run-time of 62 minutes for RCV1-v2 and 56 minutes for Jeb Bush. The run-time includes BERT fine-tuning and inference on the dataset for one category,  one further pre-training epoch setting and one active learning strategy. In contrast, LR took a mere 0.4 minutes per run. These times represent a substantial reduction from the original paper, which reported an average run-time of 18 hours (or 1,080 minutes) per single active learning strategy run. After examining their code, we discovered that the original authors used fp32 precision computing throughout their entire experiment. Initially, we followed this method but found that it largely prolonged the run-time to about 3 to 4 hours per run on our machine. Consequently, we switched to using fp16 precision, which considerably accelerated the process without noticeably affecting the results. Nonetheless, we agree with the original conclusion that when integrating BERT into the TAR workflow, its performance in terms of run-time cannot rival that of simpler baseline models like LR. 

\vspace{-10pt}
\subsection{RQ2: Does the Goldilocks Finding Generalise?}
\subsubsection{Further Pre-training.}



Next, we consider our new results on the CLEF datasets, which are out-of-domain w.r.t. to the BERT backbone. Results are reported in Tables \ref{table:4-1} and \ref{table:4-2}: the BERT-based model consistently outperforms the baseline in terms of R-Precision across almost all datasets, given the Goldilocks epoch and the corresponding active learning strategy. However, the specific pattern of the Goldilocks epoch becomes less distinct when focusing solely on R-Precision. Furthermore, it is noteworthy that the choice of active learning strategy influences the characteristics of the Goldilocks epoch. For example, CLEF 2017 train, CLEF 2018 test, and CLEF 2019 dta test exhibit identical epochs for both strategies, albeit with variations in the exact epoch number. Conversely, CLEF 2017 test and CLEF 2019 intervention test display divergent epochs under each strategy. In summary, we did not observe the 'domain mismatch' problem on CLEF TAR collections when adapting the BERT-base model to this task. We confirm that further pre-training works in TAR workflow as the performance can be improved when the number of further pre-training epochs increases among all the metrics.

Aside from investigating the Goldilocks epoch for the BERT-base model on the CLEF collections, we found that active learning strategies also affect final performance with the same epoch for further pre-training. Relevance feedback shows the tendency to identify more relevant documents during active learning and displays the least effort for screening the remaining documents after active learning has concluded. 
This is the same for review cost in the expensive training setting, as relevance feedback has the potential to rank unreviewed documents better for second-stage review with fewer reviewed documents to train. However, applying BERT requires reviewing more documents for training than the baseline LR model to obtain the same target recall, which is a significant drawback for practical applications.

\vspace{-10pt}
\subsubsection{Run-Time Analysis.} For the CLEF collections, the run-time for each topic ranges from $\approx$2.75 min (topic CD010355 with 43 documents, a rarely small number across these collections) to 42 min (topic CD009263, with 79,782 documents) for each run (20 iterations, 2 active learning strategies). Inference time and training time depend on the number of documents associated with each topic. On topics with only a few documents, the cost is dictated by the training phase; in the last iteration (thus, the one that takes the longest) this is typically up to 20 seconds. On topics with a large number of documents, the cost is dictated by the inference phase; on the largest set of documents (CD009263), it takes $\approx$49 seconds for inference per iteration, while training time is up to 17 seconds. 

\vspace{-10pt}
\subsection{RQ3: Can we Address the Domain-Mismatch Problem?}

	Since we have observed that the BERT-base model can work properly in TAR with further pre-training, we want to know what is the gap between such a paradigm and a domain-specific pre-trained model. To investigate this, we consider BioLinkBERT, the state-of-the-art BERT-like backbone across biomedical benchmarks such as BLURB and MedQA-USMLE ~\cite{yasunaga2022linkbert}. 
	Results when using the BioLinkBERT backbone with no further pre-training on the CLEF TAR datasets are reported in Tables \ref{table:4-1} and \ref{table:4-2}. We find that this domain-specific backbone can largely outperform the LR baseline and the BERT model across almost all its tested further-pre-training settings: in the CLEF 2017-2018 collections, it can achieve $\approx$0.80 R-Precision, while it achieves up to 0.93 R-Precision in CLEF 2019. For review costs, we observe a concurrent improvement when compared to the common BERT backbone. In the uniform cost setting, the BioLinkBERT backbone, without additional pre-training, uses fewer or an equal number of reviewed documents compared to LR. While it shows significant improvement under the expensive training setting, it still falls short of LR. However, considering the time spent on further pre-training with the BERT-base model, and the effort in identifying the Goldilocks epoch (which is impractical to pinpoint in real-life scenarios), choosing an appropriate model emerges as a simple and adequate solution—if such a model exists.
	These findings suggest that using a tailored pre-trained model, like the BioLinkBERT backbone in the case of the CLEF collections considered here, can improve the effectiveness of the model in the target domain, and remove the effort required in finding the Goldilocks epoch. Moreover, our results highlight the importance of selecting an appropriate pre-trained backbone for domain-specific tasks.

\begin{table}[!t]
	\def\textBF#1{\sbox\CBox{#1}\resizebox{\wd\CBox}{\ht\CBox}{\textbf{#1}}}
	\centering
	
	\caption{Results for CLEF 2017-2018 collections. \textsuperscript{$\ast$} indicates statistical significance differences w.r.t. the baseline LR.}
	\label{table:4-1}
	
	\scriptsize
	\begin{adjustbox}{width=1.0\textwidth, left=\textwidth}
		\resizebox{\textwidth}{!}{%
			\begin{tabular}{@{}cc|cc|rr|rrr@{}}
				\toprule
				\multirow{2}{*}{Collection} & \multirow{2}{*}{\begin{tabular}[c]{@{}c@{}}Further Pre-training \\ Epoch\end{tabular}} & \multicolumn{2}{c|}{R-Precision ($\uparrow$)} & \multicolumn{2}{r|}{Uni. Cost ($\downarrow$)} & \multicolumn{2}{r}{Exp. Train. ($\downarrow$)} \\
				&  & Relevance & Uncertainty & Rel. & Unc. & Rel. & Unc. \\ \midrule
				& baseline & \textBF{0.734} & \textBF{0.603} & \textBF{1.000} & \textBF{1.000} & \textBF{1.000} & \textBF{1.000} \\
				\multirow{5}{*}{CLEF 2017 train} & 0 & \textsuperscript{$\ast$}0.612 & 0.598 & 1.916 & 1.508 & \textsuperscript{$\ast$}2.233 & \textsuperscript{$\ast$}2.122 \\
				& 1 & *0.674 & 0.679 & 1.231 & 0.926 & *1.665 & *1.654 \\
				& 2 & 0.697 & 0.669 & 1.147 & 0.935 & *1.575 & *1.605 \\
				& 5 & 0.711 & 0.670 & \textBF{0.974} & 0.902 & *1.500 & *1.574 \\
				& 10 & \textBF{0.722} & \textBF{0.680} & 0.989 & \textBF{0.847} & *\textBF{1.446} & *\textBF{1.543} \\ \midrule
				& BioLinkBERT-ep0 & *\textBF{0.838} & \textBF{*0.761} & \textBF{0.727} & \textBF{0.606} & \textBF{1.174} & *\textBF{1.332} \\ \midrule
				& baseline & \textBF{0.782} & \textBF{0.657} & \textBF{1.000} & \textBF{1.000} & \textBF{1.000} & \textBF{1.000} \\
				\multirow{5}{*}{CLEF 2017 test} & 0 & 0.727 & 0.722 & *1.879 & *1.774 & *2.120 & *2.003 \\
				& 1 & 0.756 & 0.738 & *1.471 & *1.448 & *1.710 & *1.820 \\
				& 2 & 0.746 & 0.748 & *1.588 & *1.409 & *1.687 & *1.780 \\
				& 5 & \textBF{0.776} & 0.728 & *1.460 & *1.561 & \textBF{*1.578} & *1.756 \\
				& 10 & 0.776 & \textBF{0.762} & \textBF{1.424} & \textBF{1.331} & *1.601 & \textBF{*1.750} \\ \midrule
				& BioLinkBERT-ep0 & \textBF{0.812} & \textBF{*0.794} & \textBF{1.016} & \textBF{1.058} & \textBF{*1.545} & \textBF{*1.645} \\ \midrule
				& baseline & \textBF{0.754} & \textBF{0.644} & \textBF{1.000} & \textBF{1.000} & \textBF{1.000} & \textBF{1.000} \\
				\multirow{5}{*}{CLEF 2018 test} & 0 & 0.694 & 0.695 & 1.655 & 1.863 & *2.174 & *2.280 \\
				& 1 & 0.725 & 0.725 & 1.485 & 1.617 & *1.947 & 2.115 \\
				& 2 & 0.720 & 0.722 & 1.246 & 1.776 & *1.674 & 2.158 \\
				& 5 & 0.729 & 0.729 & 1.148 & 1.454 & *1.668 & *1.934 \\
				& 10 & \textBF{0.747} & \textBF{0.750} & \textBF{1.067} & \textBF{1.021} & \textBF{*1.596} & \textBF{*1.585} \\ \midrule
				& BioLinkBERT-ep0 & \textBF{0.793} & \textBF{*0.780} & \textBF{0.669} & \textBF{0.801} & \textBF{1.222} & *\textBF{1.436} \\ \bottomrule
			\end{tabular}%
		}
	\end{adjustbox}
\vspace{-15pt}
\end{table}

\begin{table}[!t]
	\def\textBF#1{\sbox\CBox{#1}\resizebox{\wd\CBox}{\ht\CBox}{\textbf{#1}}}
	\centering
		\caption{Results for CLEF 2019 collections. \textsuperscript{$\ast$} indicates statistical significance differences w.r.t. the baseline LR.
		}
	\label{table:4-2}
	\scriptsize
	\begin{adjustbox}{width=1.0\textwidth, left=\textwidth}
		\resizebox{\textwidth}{!}{%
		\begin{tabular}{@{}cc|cc|cc|cc@{}}
			\toprule
			\multirow{2}{*}{Collection} & \multirow{2}{*}{\begin{tabular}[c]{@{}c@{}}Further Pre-training \\ Epoch\end{tabular}} & \multicolumn{2}{c|}{R-Precision ($\uparrow$)} & \multicolumn{2}{c|}{Uni. Cost ($\downarrow$)} & \multicolumn{2}{c}{Exp. Train. ($\downarrow$)} \\
			&  &  \multicolumn{1}{c}{Relevance} & Uncertainty & Rel. & Unc. & Rel. & Unc. \\ \midrule
			& baseline & \textBF{0.823} & \textBF{0.742} & \textBF{1.000} & \textBF{1.000} & \textBF{1.000} & \textBF{1.000} \\
			\multirow{5}{*}{\begin{tabular}[c]{@{}c@{}}CLEF 2019 \\ dta test\end{tabular}} & 0 & 0.775 & 0.803 & 1.856 & 1.795 & 3.549 & 3.520 \\
			& 1 & 0.802 & 0.793 & 1.787 & 1.736 & 3.349 & 3.573 \\
			& 2 & 0.804 & 0.798 & 2.144 & 1.709 & 3.508 & \textBF{3.279} \\
			& 5 & \textBF{0.838} & \textBF{0.833} & 1.788 & 1.820 & \textBF{2.993} & 3.352 \\
			& 10 & 0.791 & 0.818 & \textBF{1.523} & \textBF{1.542} & 3.020 & 3.444 \\ \midrule
			& BioLinkBERT-ep0 & \textBF{0.909} & \textBF{0.857} & \textBF{0.979} & \textBF{1.109} & *\textBF{2.188}  &\textBF{*2.555} \\ \midrule
			& baseline & \textBF{0.913} & \textBF{0.843} & \textBF{1.000} & \textBF{1.000} & \textBF{1.000} & \textBF{1.000} \\
			\multirow{5}{*}{\begin{tabular}[c]{@{}c@{}}CLEF 2019 \\ intervention \\ train\end{tabular}} & 0 & 0.898 & 0.891 & 1.206 & 1.254 & *1.908 & *1.834 \\
			& 1 & 0.903 & 0.893 & 1.068 & 1.311 & *1.997 & *1.892 \\
			& 2 & 0.912 & 0.919 & \textBF{1.015} & 0.976 & *1.833 & \textBF{*1.834} \\
			& 5 & \textBF{0.924} & 0.920 & 1.029 & 0.950 & \textBF{*1.816} & *1.851 \\
			& 10 & 0.913 & \textBF{0.929} & 1.079 & \textBF{0.934} & *1.853 & *1.869 \\ \midrule
			& BioLinkBERT-ep0 & \textBF{0.939} & 0.923 & \textBF{0.902} & \textBF{0.620} & \textBF{*1.596} & \textBF{*1.617} \\ \midrule
			& baseline & \textBF{0.855} & \textBF{0.759} & \textBF{1.000} & \textBF{1.000} & \textBF{1.000} & \textBF{1.000} \\
			\multirow{5}{*}{\begin{tabular}[c]{@{}c@{}}CLEF 2019 \\ intervention\\ test\end{tabular}} & 0 & 0.813 & 0.785 & *1.678 & *1.417 & *2.486 & *2.457 \\
			& 1 & 0.851 & 0.813 & *1.370 & 1.272 & *2.185 & *2.264 \\
			& 2 & 0.859 & \textBF{0.860} & *1.308 & 1.089 & *2.131 & *2.275 \\
			& 5 & \textBF{0.872} & 0.852 & \textBF{1.247} & \textBF{1.060} & *2.030 & *\textBF{2.224} \\
			& 10 & 0.820 & 0.799 & 1.342 & 1.285 & *\textBF{2.027} & *2.312 \\ \midrule
			& BioLinkBERT-ep0 & \textBF{0.934} & *\textBF{0.900} & \textBF{0.869} & \textBF{0.773} & \textBF{*1.748} & \textBF{*1.967} \\ \bottomrule
		\end{tabular}%
	}
  \end{adjustbox}
\vspace{-22pt}
\end{table}


	\label{fig:3}

\section{Discussion}
\vspace{-10pt}
In our study, we experimented with both original and domain-specific BERT-based models (i.e. BERT-base and BioLinkBERT) as the backbone of the active learning pipeline for TAR. BioLinkBERT is a BERT model pre-trained on the PubMed abstract corpus with additional citation links. While it includes the standard MLM, BioLinkBERT replaces the next sentence prediction (NSP) task with the document relation prediction (DPR) task. DRP is specifically designed to understand the relationships between documents, which could involve linking concepts. In our experiment with the domain-specific CLEF TAR collections, BioLinkBERT outperformed the BERT-base model without requiring any further pre-training. It also surpassed the baseline logistic regression (LR), which the BERT-base model may not outperform, even at its Goldilocks epoch.

Recently, models in the DeBERTa series, especially DeBERTa-v3 \cite{he2021debertav3}, have demonstrated superior performance compared to the BERT model in classification tasks. DeBERTa-v3 introduces a key modification by replacing MLM with replaced token detection (RTD), an ELECTRA-style pre-training task \cite{clark2020electra} aims to train a discriminator to distinguish whether a token in the text has been replaced by the generator. We experimented with both DeBERTa-base \cite{he2020deberta} and DeBERTa-v3-base models as the backbone for the CLEF TAR collections. However, these did not show more promising results compared to BERT-like backbones. The results will be updated in our GitHub repository.

\section{Conclusions}
\vspace{-10pt}


In this paper, we investigated an active learning pipeline for TAR based on BERT models through the reproduction of previously published experiments, and extension to a new TAR context, that of screening of studies for the creation of medical systematic reviews. 

Our reproduction of the original study was not fully successful: we were not able to obtain the same results, which we suggest are largely dataset-specific and pre-processing-specific.
For instance, the optimal number of further pre-training epochs (i.e. the Goldilocks) in our experiments differs from that of the original paper. However, our reproduction confirms the general findings of the original paper: that while there exists a Goldilocks epoch that is best used for further pre-training BERT-based models for TAR in an active learning setting, this epoch is hard to determine a priori and it largely depends on the dataset.


Our study further sheds light on the 'domain-mismatch' problem highlighted by the original work. Despite the notable conceptual differences between the CLEF collections and the pre-training corpora used to obtain BERT, we show that the BERT base model can still outperform the baseline in these collections when the Goldilocks epoch is identified. Additionally, our experiments showcase the effectiveness of a domain-specific pre-trained model, BioLinkBERT, in the TAR workflow when medical systematic reviews are considered. Remarkably, this model surpasses both BERT set to the Goldilocks epoch and the logistic regression baseline, without the need for any further pre-training, thus saving a consistent amount of additional computation.

Our results also resonated with previous studies \cite{lewis1995sequential,cormack2014evaluation} on the influences of active learning strategies: relevance feedback can benefit logistic regression more than uncertainty sampling. However, when using BERT for TAR, the choice of active learning strategy is not consistent across metrics, particularly on the RCV1-v2 and Jeb Bush datasets. Interestingly, on the CLEF-TAR collections, instead, relevance feedback consistently yields better results on R-Precision and the expensive training cost. 


Overall, our findings suggest that the search for the Goldilocks epoch is a laborious way of improving the effectiveness of BERT-based classifier models in TAR. Instead, we suggest that considering the task's characteristics and identifying an appropriate pre-trained BERT-like backbone may be a simpler and more effective way to achieve better effectiveness in TAR tasks. We also notice a recent study  \cite{molinari2022transferring} has shown these pre-trained models can further improve the active learning process with zero-shot rankings, which indicates a promising trend in their application on professional retrieval tasks such as systematic review screening.

\section*{Acknowledgment}
Xinyu Mao is supported by a UQ Earmarked PhD Scholarship and this research is funded by the Australian Research Council Discovery Projects programme ARC DP DP210104043.

\clearpage
	
\bibliographystyle{splncs04}
\bibliography{bibliography}


\end{document}